\documentclass[aps,pra,reprint,showpacs,superscriptaddress]{revtex4-1}
\usepackage{epsfig}
\usepackage{natbib,color,epstopdf}
\usepackage{amsfonts}
\usepackage{graphicx}
\usepackage{amsmath, amsthm}
\usepackage{bm}
\usepackage{color}
\newcommand{\be}{\begin{equation}}
\newcommand{\ee}{\end{equation}}
\newcommand{\bwt}{\begin{widetext}}
\newcommand{\ewt}{\end{widetext}}
\newcommand{\bea}{\begin{eqnarray}}
\newcommand{\eea}{\end{eqnarray}}
\newcommand{\ket}[1]{|#1\rangle}
\newcommand{\bra}[1]{\langle #1|}

\newcommand{\ip}[2]{\langle #1|#2 \rangle}

\usepackage{lipsum}

\begin{document}
\title{Equation of Motion for Estimation Fidelity of Monitored Oscillating Qubits}
\author{Humairah Bassa}
\affiliation{School of Chemistry and Physics, University of KwaZulu, Natal, Durban, South Africa}
\author{Lajos \surname{Di\'osi}}
\affiliation{ Wigner Research Centre for Physics, Institute for Particle and Nuclear Physics, H-1525 Budapest 114, P.O.B. 49, Hungary}
\author{Thomas Konrad}
\email{konradt@ukzn.ac.za}
\affiliation{School of Physics, University of KwaZulu, Natal, Durban, South Africa}
\affiliation{National Institute of Theoretical Physics, UKZN Node}
\author{Hermann Uys}
\email{hermann.uys@gmail.com}
\affiliation{National Laser Centre, Council for Scientific and Industrial Research, Pretoria, South Africa}
\affiliation{Department of Physics, Stellenbosch University, Stellenbosch, South Africa}
\begin{abstract}

We study the convergence properties of state estimates of an oscillating qubit being monitored by a sequence of \textit{discrete}, unsharp measurements. Our method derives a differential equation determining the evolution of the estimation fidelity from a single incremental step. When the oscillation frequency $\Omega$ is precisely known, the estimation fidelity  converges exponentially fast to unity.  For imprecise knowledge of  $\Omega$ we derive the asypmtotic estimation fidelity.          
\end{abstract}

\maketitle
\section{Introduction.}

High fidelity quantum state estimation is a key requirement in innumerable quantum control applications of quantum information processing, quantum simulation, quantum metrology, and quantum conmmunication.  Quantum state estimation \cite{Doherty.et.al00, Diosi2006, Oxtoby2008, KonradUys2012} based on continuous or sequential unsharp (sometimes called weak) measurement \cite{Diosi1988, Belavkin1989, Wiseman1993, Carmichael93, Korotkov2001, Audretsch2001} has opened new avenues for quantum control that obviate the need for repeated state preparation to execute tomography and allow, for example, real-time quantum closed-loop feedback.   These principles have been brought to bear in different experimental platforms including microwave cavities \cite{Sayrin2011} and superconducting qubits \cite{Vijay2012}. Improved sophistication in the unsharp measurement control toolbox \cite{Vool2016} promises significant expansion beyond traditional open loop quantum control applications.

To achieve high fidelity control based on unsharp measurement the experimenter is forced to balance the benefit of allowing coherent dynamics to proceed subject to only weak perturbations, with the price of reduced information gain per measurement. As such, finding optimal estimation and control strategies are of prime importance.  To make headway, detailed analytical descriptions of the measurement and estimation process are desirable. 

In this paper we study the dynamics of  state estimation fidelity during a sequence of discrete, unsharp  measurements.  Detailed analytical results are natural in the domain of \emph{continuous} unsharp measurement, but we attempt here to place on a firmer footing the understanding of estimation dynamics during \textit{sequential,  discrete } measurements, as is natural in many experimental settings like trapped ions or microwave cavities.  

We consider an estimation protocol wherein a state estimate is  propagated by a Hamiltonian presumed to drive a laboratory quantum system which is also subject to sequential unsharp measurement.  The state estimate is sequentially updated  based on the outcome of each measurement on the actual quantum state with the same propagator as the system. Numerical simulations have demonstrated the convergence of the state estimate when all parameters in the Hamiltonian are precisely known~\citep{Rothe2010, KonradUys2012} and  even for the case of process tomography, treating the Hamiltonian parameters as state variables of a hybrid system \cite{Ralph2011, Molmer2013, Bassa2015}.

Here we study the convergence of the state estimate for sequential unsharp measurements analytically, for the case of a two-level system undergoing Rabi oscillations. We start our study in Section~\ref{Section:fidelity} with a brief explanation of the state estimation protocol.  In Section~\ref{Section:Rabi} we investigate the dynamics of the estimation fidelity, when the Hamiltonian is precisely known, or known within a specified error margin. This allows us to place bounds on the parameter space that grants high-fidelity state estimation. 

\section{State estimation and fidelity}\label{Section:fidelity}

A commonly used distance measure for the ``closeness'' of two quantum states is the estimation fidelity. It is defined as $F(\rho,\rho_e)=\mathrm{Tr}\left[\sqrt{\rho^\frac{1}{2}\rho_e\rho^{\frac{1}{2}}}\right]^2$, and is sometimes referred to as the \textit{squared fidelity}. Here $\rho$ and $\rho_e$ are the density matrices describing the actual quantum state and the estimate of that quantum state, respectively.  When both states are pure, i.e. $\rho = \ket{\psi}\bra{\psi}$ and $\rho_e=\ket{\psi^e}\bra{\psi^e}$, the fidelity takes the simple form $F(\psi,\psi^e)=\vert\langle{\psi}\ket{\psi^e}\vert^2$ \cite{Jozsa1994}.  The greater the fidelity, the more similar the two states are. It is $0$ if, and only if,  $\ket{\psi}$ and $\ket{\psi^e}$ are orthogonal pure states and it is $1$ if, and only if,  $\ket{\psi}=\ket{\psi^e}$. 

Here we assume that $\ket{\psi}$ and $\ket{\psi^e}$ represent an \emph{actual} and an \emph{estimated} state, respectively. These states do not in general initially coincide, and our task is to exploit unsharp measurement with the aim of forcing $\ket{\psi^e}$ to converge onto $\ket{\psi}$ in the presence of ongoing dynamics. A single elementary step of the estimation protocol consists of a unitary time evolution of both states followed by an unsharp measurement (which is a probabilistic  ``filtering'' operation) on the actual state. The estimated state is updated based on the result of the measurement on the actual system. Under appropriate circumstances successive repetitions of this elementary step leads to a faithful estimate of the actual state in real time.

More concretely, an elementary step of the method can be formulated mathematically as follows. First, the actual state evolves according to the Hamiltonian dynamics of the system specified by the evolution operator $\hat U(\Omega)$:
\begin{equation}
    \ket{\psi}\Rightarrow \hat U(\Omega)\ket \psi \equiv\ \ket{\psi'}.
\end{equation}
If the Hamiltonian is precisely known then the estimate is propagated using the same evolution operator. However, a (classical) parameter $\Omega$ specified in the Hamiltonian, such as the Rabi frequency in the case of Rabi oscillations, may be detuned away from the actual parameter. The estimated state is thus evolved using an estimated unitary operator $U(\Omega_e)$:
\begin{equation}
    \ket{\psi^e}\Rightarrow\hat  U(\Omega_e)\ket{\psi^e} \equiv\ \ket{\psi^{e'}}.
\end{equation}
The task of estimating $\Omega_e$ has been the subject of earlier work  \cite{Bassa2015}. The actual state then undergoes the following random change (collapse) under selective measurement,
\begin{equation}\label{selmeas}
    \ket{\psi'}\Rightarrow\frac{1}{\sqrt{p_n}}\hat M_n\ket{\psi'}\equiv\ket{\psi'_{n}}
\end{equation}
where $\hat M_n$ is the Kraus operator corresponding to the  n'th allowed measurement outcome, and $p_n=\bra{\psi} \hat M_n^\dagger\hat  M_n\ket{\psi}$ is the associated probability for the outcome. The Kraus operators are constrained via $\sum_n \hat M_n^\dagger\hat M_n=\sum_n\hat  E_n=\mathbb{I}$, where we introduced the positive effects $\hat E_n=\hat M_n^\dagger \hat M_n$. The estimate $\ket{\psi^e}$ is updated using the outcome of the selective measurement just done on $\ket{\psi'}$:
\begin{equation}\label{selupd}
    \ket{\psi^e}\Rightarrow\frac{1}{\sqrt{p^e_n}}\hat M_n\ket{\psi^{e'}}\equiv\ket{\psi^{e'}_{n}}
\end{equation}
with $p^e_n=\bra{\psi^{e'}} \hat E_n\ket{\psi^{e'}}$. The divisor $p^e_n$ is merely used to re-normalize the updated $\ket{\psi^{e'}}$; the statistics of the updates are determined by the probabilities $p_n$ to observe outcome $n$.

We can now define the \emph{average} change in fidelity after a single elementary step of the estimation protocol:
\begin{equation}\label{eq:DeltaF}
    \Delta F = \sum_n p_n \vert\ip{\psi'_n}{\psi^{e'}_n}\vert^2~-~\vert\ip{\psi}{\psi^e}\vert^2.
\end{equation}
Using Eq.~\eqref{selmeas} and~\eqref{selupd} we obtain
\begin{equation}\label{changeFidbefore}
    \Delta F = \sum_n \frac{\vert\bra{\psi}\hat U^\dagger(\delta)\hat E'_n\ket{\psi^e}\vert^2}{\bra{\psi^e}\hat E'_n\ket{\psi^e}}~-~\vert\ip{\psi}{\psi^e}\vert^2,
\end{equation}
where $\delta =\Omega-\Omega_e$ and $\hat E'_n=\hat U^\dagger(\Omega_e)\hat E_n\hat U(\Omega_e)$ are the time-varying effects.

As is clear from Ref.~\cite{Diosi2006}, in the case where the Hamiltonian is precisely known ($\delta=0$) the above update follows the spirit of  the classical Bayesian update and we expect intuitively that the measured actual $\ket{\psi}$ and the updated estimate $\ket{\psi^e}$ come ``closer'' to each other. The method performs suitably well in various quantum estimation situations \cite{Rothe2010, KonradUys2012, Sayrin2011, Hiller2012}. In what follows, we will use Eq.~\eqref{changeFidbefore}  to derive time dependence of the estimation fidelity convergence of oscillating qubits when the Hamiltonian is precisely know ($\delta=0$) and find the asymptotic estimation fidelity when the Hamiltonian is not precisely known ($\delta \not= 0$).

\section{ Rabi oscillations}\label{Section:Rabi}

Consider a single two-level system undergoing Rabi oscillations due to the Hamiltonian
\begin{equation}
  \hat{H}(\Omega)= \frac{\Omega}{2}\hat\sigma_x\label{RabiH}
\end{equation}
where $\hbar=1$, $\Omega$ is the Rabi frequency and $\hat\sigma_x$ is the Pauli matrix that generates rotations about the $x$-axis. The corresponding evolution operator is
\begin{equation}
  \hat{U}(\Omega) = \exp(-i\hat{H}(\Omega)\tau).
\end{equation}
In order to estimate the state of the system we perform symmetric unsharp measurements of the $\hat\sigma_z$ observable~\cite{KonradUys2012}. The corresponding effects are given by
\begin{eqnarray}
    E_0 = \tfrac{1}{2}(\mathbb{I}+\Delta p \hat\sigma_z) \\
    E_1 = \tfrac{1}{2}(\mathbb{I}-\Delta p \hat\sigma_z)
\end{eqnarray}
where $\Delta p$ is the strength of the individual measurements. The time-varying effects are then
\begin{eqnarray}
    E'_0 = \tfrac{1}{2}[\mathbb{I}+\Delta p(\cos(\Omega_e\tau)\hat\sigma_z+\sin(\Omega_e\tau)\hat\sigma_y)] \\
    E'_1 = \tfrac{1}{2}[\mathbb{I}-\Delta p(\cos(\Omega_e\tau)\hat\sigma_z+\sin(\Omega_e\tau)\hat\sigma_y)].
\end{eqnarray}

\subsection{Incremental equation}
We find that the change in fidelity (Eq.~\eqref{changeFidbefore}) reduces to
\begin{widetext}
\begin{equation}\label{eq:RabiChangeF}
  \Delta F = \frac{\Delta p^2\left[\sin^2\left(\Omega_e\tau+\theta_e\right)\sin^2\left(\tfrac{\theta-\theta_e}{2}\right)\right]-(1-\Delta p^2)\left[\cos^2\left(\tfrac{\theta-\theta_e}{2}\right)-\cos^2\left(\tfrac{\theta-\theta_e}{2}+\tfrac{\delta\tau}{2}\right)\right]}{1-\Delta p^2 \cos^2\left(\Omega_e\tau+\theta_e\right)}
\end{equation}
\end{widetext}
where $\theta,\theta_e \in [0,\pi]$ are the polar angles of the Bloch vectors corresponding to the states $\ket{\psi}$ and $\ket{\psi^e}$, respectively. The first term in Eq.~\eqref{eq:RabiChangeF} is the change in fidelity due to the measurement (which is performed after the unitary evolution), and is modulated by the measurement strength.  For a projective measurement ($\Delta p ^2 = 1$), we see that the change in fidelity is maximal, i.e. unity minus the initial fidelity, $\cos^2(\tfrac{\theta-\theta_e}{2})$. The second term is the change in fidelity due to the evolution. It does not contribute to the overal change in fidelity in two cases -- when the Rabi frequency is precisely known, as well as when the measurement is projective.  We notice that the right-hand side of the equation is a convex sum -- the stronger the measurement, the weaker the influence of the unitary dynamics. In other words, if the level-resolution time  $T_{\mathrm{lr}}= \tau/\Delta p ^2$ , which defines the timescale on which the state evolves due to the measurement sequence,  is much briefer than the timescale $1/\delta$ of the unitary dynamics,  
i.e.  $\tau \delta/(\Delta p^2) \ll 1$, then the measurement has a greater influence on the change in fidelity than the relative dynamics between the state and the estimate.


\subsection{Differential equation}

Hitherto we have restricted ourselves to a formalism appropriate for describing a sequence of discrete, unsharp measurements. Specifically, we've refrained from using stochastic Schrodinger equation or Ito calculus approaches suited to continuous measurement scenarios.  To obtain analytical expressions governing the ensemble averaged behaviour of our protocol we are ultimately forced to take a time continuous limit.  In this way we can derive a differential equation for the average change in fidelity after a single elementary step of the estimation protocol. 
\begin{figure}[h!]
\centering
\includegraphics[width= \columnwidth,keepaspectratio]{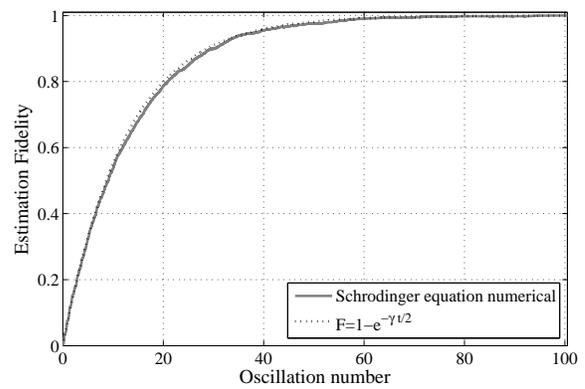}
\caption{Convergence of the estimation fidelity for known dynamics.  The solid grey line shows the averaged fidelity over 1000 numerical simulations , while the dotted line represents the theoretical prediciton Eq.~(\ref{exponential}).  We used $\Delta p= 0.04$ and $\tau=\pi/50$ so that $\gamma=0.0255$, and averaged over 1000 runs. }\label{expfig}
\end{figure}

To this end we make two changes to ~Eq.~\eqref{eq:RabiChangeF}.  Firstly, we transform to a new coordinate system for the polar angles of the two Bloch vectors by defining, respectively, the relative half-angle, $\theta_r = (\theta-\theta_e)/2$, and the mean angle, $\bar\theta=(\theta+\theta_e)/2$.  In addition, we can rewrite the equation in terms of the fidelity since it relates to the relative angle via  $F=\cos^2{\theta_r}$. This yields
\begin{widetext}
\begin{equation}
	\Delta F = \frac{\Delta p^2\sin^2\left(\Omega_e\tau+\bar\theta-\theta_r\right)(1-F)-(1-\Delta p^2)[\pm\sqrt{F(1-F)}\sin(\delta\tau)+(F-\tfrac{1}{2})(1-\cos(\delta\tau))]}{1-\Delta p^2 \cos^2\left(\Omega_e\tau+\bar\theta-\theta_r\right)}.\label{dFmeanangle}
\end{equation}
\end{widetext}
Note that the two possible signs in front of the square root come about when relating a cosine and sine product to the fidelity since $\cos{\theta_r}\sin{\theta_r}=\sqrt{F(1-F)}$ or  $\cos{\theta_r}\sin{\theta_r}=-\sqrt{F(1-F)}$ depending on the particular value of $\theta_r$. In the regime  where the sequential measurement strength is weak compared to the unitary dynamics, the fidelity will change little over the course of a single Rabi oscillation.   Therefore, we can average over all possible mean angles using $\overline{\sin^2\bar\theta}=\overline{\cos^2\bar\theta}=1/2$ and $\overline{\sin\bar\theta}=\overline{\cos\bar\theta}=0$, where the overline indicates this average over oscillations. This average is taken after trigonometrically expanding the sine in the first term in the numerator and the cosine in the denominator.  In order to obtain a differential equation we divide by the change in time after a single elementary step, $\tau$,  and take the limit that $\tau$ tends to zero. Simultaneously we  require that $\Delta p$ tends to zero such that $\lim_{\substack{\Delta p \rightarrow 0\\\tau\rightarrow 0}}\Delta p^2/\tau =\gamma$, the so-called continuum limit of the sequence of unsharp measurements \cite{AudretschDiosiKonrad02}. We thus arrive at: 
\begin{equation}\label{eq:differential}
	\frac{\mathrm{d}F}{\mathrm{d}t}=\frac{\gamma}{2}(1-F)\pm\delta\sqrt{F(1-F)}.
\end{equation}
The parameter $\gamma$ characterizes the strength of the measurement sequence \cite{AudretschDiosiKonrad02} and is related to the level resolution time $T_{\mathrm{lr}}= 1/\gamma $. Implicit in this derivation is the assumption $\delta\tau~\ll~1$. 

If the Rabi frequency is known precisely (i.e. $\delta=0$), this equation has the simple solution
\begin{equation}\label{exponential}
  F = 1- \exp\left(-\tfrac{\gamma}{2}t\right),
\end{equation}
The average estimation fidelity of a sequential unsharp measurement thus converges exponentially fast to unity. This result was already conjectured from numerical simulations in~\cite{KonradUys2012}.  Fig.~\ref{expfig} compares the theoretical prediction (\ref{exponential})  to the averaged estimation fidelity obtained from   
1000 numerical simulations of a qubit undergoing Rabi oscillations governed by Hamiltonian (\ref{RabiH}), while being subjected to unsharp measurements at periodicity $\tau$. We used $\Delta p= 0.04$ and $\tau=\pi/50$, yielding $\gamma=0.0255$ in units of the Rabi frequency $\Omega$. 

\begin{figure}[h!]
\centering
\includegraphics[width= \columnwidth,keepaspectratio]{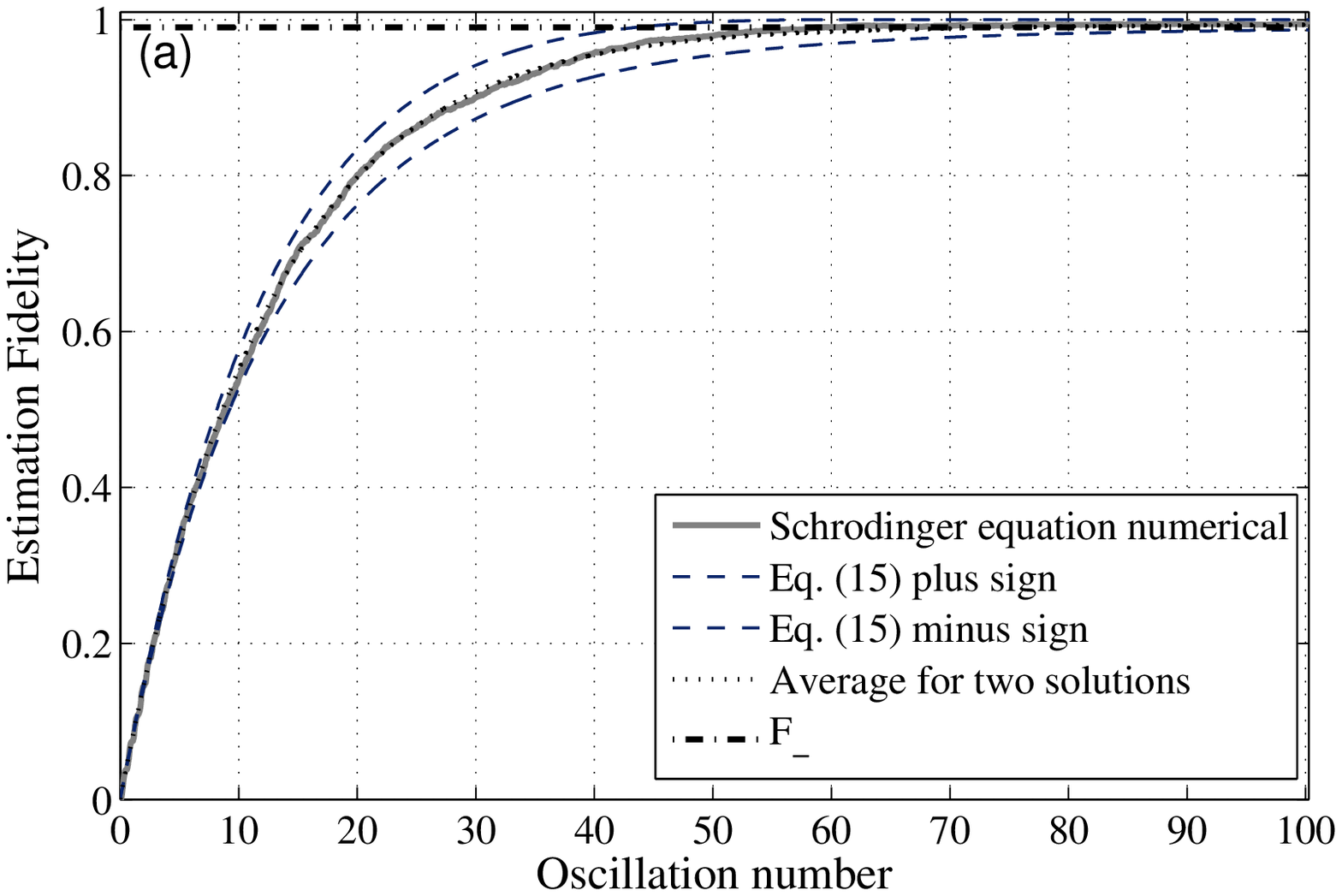}
\includegraphics[width= \columnwidth,keepaspectratio]{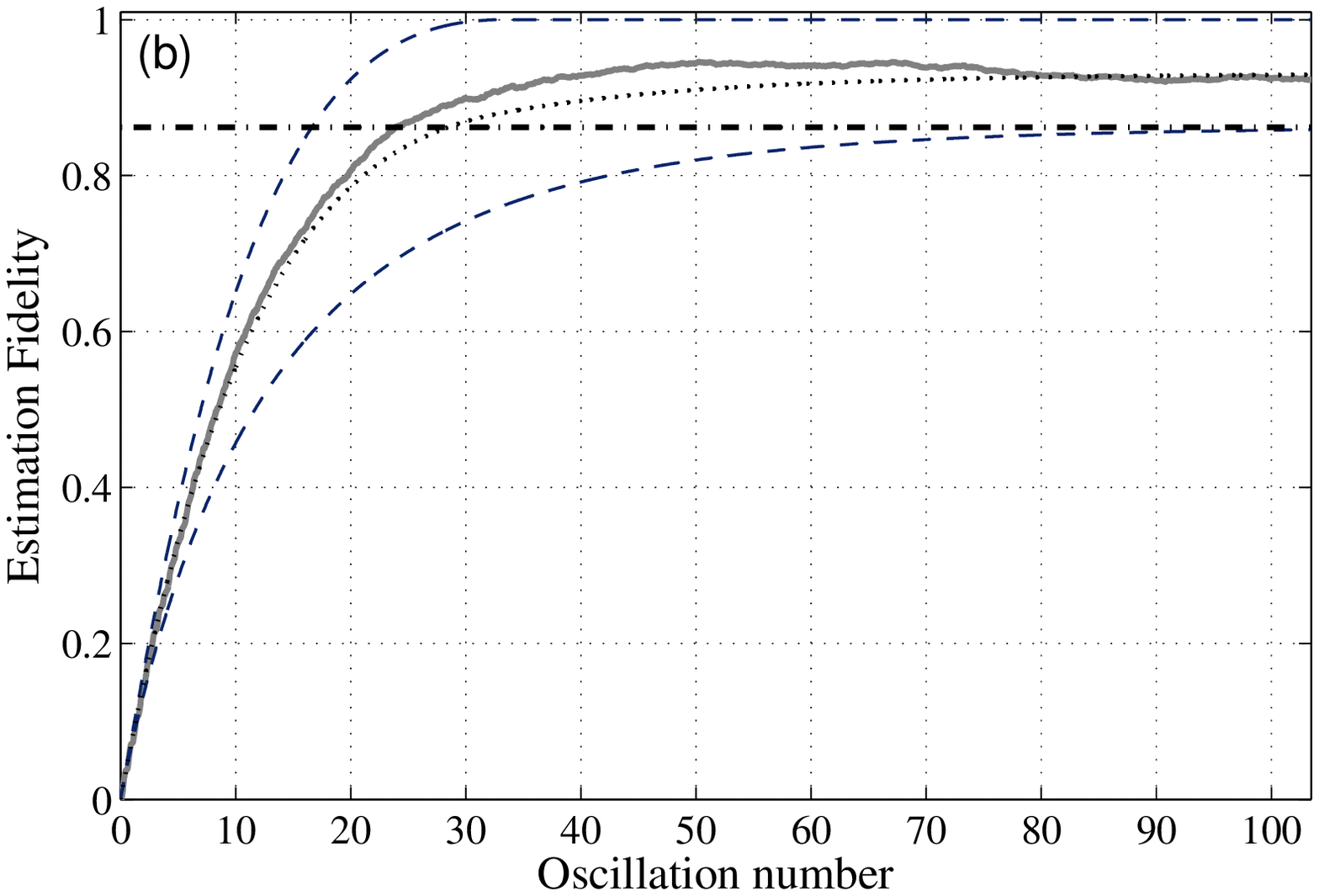}
\caption{Convergence of the estimation fidelity for finite $\delta$. The solid grey line shows the numerical solution to the Sch\"odinger equation, the top and bottom dashed lines the numerical solutions to Eq.~\eqref{eq:differential} for the plus and minus signs respectively, while the dotted line is the average of the latter two curves. The horizontal dot-dash line is the asymptotic limit $F_-$.  In (a) $\delta=\gamma/20$ and in (b) $\delta=\gamma/5$.  In both figures $\Delta p= 0.04$ and $\tau=\pi/50$ so that $\gamma=0.0255$. }\label{Fig:delfig}
\end{figure}


If the estimate of the Rabi frequency is different from the actual value, i.e. $\delta\neq0$, the fidelity of estimation will erode. Solving Eq.~\eqref{eq:differential} at steady state, $\frac{\mathrm{d}F}{\mathrm{d}t}=0$ , we find that the asymptotic average fidelity is
\begin{equation}
	F_+=1\;\;\;\;\; \textnormal{or}\;\;\;\;\; F_- = \frac{\gamma^2}{\gamma^2+4\delta^2},
\end{equation} 
where the subscripts indicate the asymptotic solutions for the corresponding signs of the two possibilities in Eq.~\eqref{eq:differential}. The ensemble averaged behaviour is expected to be  simply the average of the two solutions to Eq.~\eqref{eq:differential} tending asymptotically to $\bar F = (F_++F_-)/2$.  This expectation is clearly borne out as illustrated in Fig.~\ref{Fig:delfig} where we use the same parameters as in Fig.~\ref{expfig}.  In Fig.~\ref{Fig:delfig}(a) $\delta=\gamma/20$, while in (b) $\delta=\gamma/5$.   The solid grey line again is the numerical solution to the Schr\"odinger equation in the presence of unsharp measurement. The dashed curves show the two possible solutions of differential equation \eqref{eq:differential} corresponding to the two allowed signs.  The dotted curve, which closely follows the numerical solution, is the average of these two solutions.  The horizontal dot-dashed line indicates the asymptotic limit $F_-$.   The slight overshoot in the numerical simulation compared to the analytical result for larger $\delta$ around the time $\sim 1/\gamma$, is persistent in our simulations and not accounted for in this model. However, very faithful correspondence between our analytical results and the numerical simulation is observed in the asymptotic regime.

\section{Conclusion}
The techniques employed in this manuscript establish a systematic approach toward analytical description of an oscillating qubit undergoing sequential, discrete unsharp measurement and can in principle be extended to other systems of interest.  Employing the same approach is expected to yield detailed understanding of process- and parameter estimation dynamics.   Our results clearly delineate parameter regimes in which sequential unsharp measurement can be usefully employed as a state estimation tool for oscillating qubits.

\section{Acknowledgement}
We dedicate this work to the memory of Ms.~Humairah Bassa whom contributed majour portions of the analytical derivations and the initial drafting of the manuscript.  We are saddened by her untimely passing away on 29 November 2015.  She was a rising star of quantum physics, a respected role model to young Islamic women such as herself, and a beloved student, colleague and friend. 

The work in this paper was supported in part by the National Research Foundation of South Africa through grant no.~93602 as well as an award by the United States Airforce Office of Scientific Research, award no.~ FA9550-14-1-0151.

\bibliographystyle{unsrt}
\bibliography{xbib}{}

\begin{thebibliography}{10}

\bibitem{Doherty.et.al00}
A.C. Doherty, S.M. Tan, A.S. Parkins, and D.F. Walls.
\newblock State determination in continuous measurement.
\newblock {\em Phys.\ Rev.}, A 60:2380, 1999.

\bibitem{Diosi2006}
L.~Di\'{o}si, T.~Konrad, A.~Scherer, and J.~Audretsch.
\newblock Coupled ito equations of continuous quantum state measurement and
  estimation.
\newblock {\em J. Phys A}, 39:L575--L581, 2006.

\bibitem{Oxtoby2008}
N.P. Oxtoby, J.~Gambetta, and H.M. Wiseman.
\newblock Model for monitoring of a charge qubit using a radio-frequency
  quantum point contact including experimental imperfections.
\newblock {\em Phys.\ Rev.}, B 77:125304, 2008.

\bibitem{KonradUys2012}
T.~Konrad and H.~Uys.
\newblock Maintaining quantum coherence in the presence of noise through state
  monitoring.
\newblock {\em Phys. Rev. A.}, 85:012102, 2012.

\bibitem{Diosi1988}
L.~Di\'{o}si.
\newblock Continuous quantum measurement and ito-formalism.
\newblock {\em Phys. Lett. A}, 129:419--423, 1988.

\bibitem{Belavkin1989}
V.~P. Belavkin.
\newblock Nondemolition measurement, nonlinear filtering, and dynamic
  programming of quantum stochastic process.
\newblock In Austin Blaqui\'{e}re, editor, {\em Modeling and Control of
  Systems}, volume 121 of {\em Lecture Notes in Control and Information
  Sciences}, pages 245--265. Springer Berlin Heidelberg, 1989.

\bibitem{Wiseman1993}
H.~M. Wiseman and G.~J. Milburn.
\newblock Quantum theory of field-quadrature measurements.
\newblock {\em Phys. Rev. A}, 47:642, 1993.

\bibitem{Carmichael93}
H.~Carmichael.
\newblock {\em An Open Systems Approach to Quantum Optics}.
\newblock Springer Verlag, Berlin, 1993.

\bibitem{Korotkov2001}
A.~N. Korotkov.
\newblock Selective quantum evolution of a qubit state due to continuous
  quantum measurement.
\newblock {\em Phys. Rev. B}, 63:115403, 2001.

\bibitem{Audretsch2001}
J.~Audretsch, T.~Konrad, and A.~Scherer.
\newblock A sequence of unsharp measurements enabling a real-time visualization
  of a quantum oscillation.
\newblock {\em Phys. Rev. A}, 63:052102, 2001.

\bibitem{Sayrin2011}
C.~Sayrin~\textit{et al}.
\newblock Real-time quantum feedback prepares and stabilizes photon number
  states.
\newblock {\em Nature}, 477:73, 2011.

\bibitem{Vijay2012}
R.~Vijay, C.~Macklin, D.H. Slichter, S.J. Weber, K.W. Murch, R.~Naik, A.N.
  Korotkov, and I.~Siddiqi.
\newblock Stabilizing rabi oscillations in a superconducting qubit using
  quantum feedback.
\newblock {\em Nature}, 490:77--80, 2012.

\bibitem{Vool2016}
U.~Vool~\textit{et al}.
\newblock Continuous quantum nondemolition measurement of the transverse
  component of a qubit.
\newblock {\em Phys. Rev. Lett.}, 117:133601, 2016.

\bibitem{Rothe2010}
T.~Konrad, Rothe A., F.~Petruccione, and L.~Di\'{o}si.
\newblock Monitoring the wave function by time continuous position measurement.
\newblock {\em New Journal of Physics}, 12:043038, 2010.

\bibitem{Ralph2011}
J.~F. Ralph, K.~Jacobs, and C.~D. Hill.
\newblock Frequency tracking and parameter estimation for robust quantum state
  estimation.
\newblock {\em Phys. Rev. A}, 84:052119, 2011.

\bibitem{Molmer2013}
A.~Negretti and K.~Molmer.
\newblock Estimation of classical parameters via continuous probing of
  complementary quantum observables.
\newblock {\em New J Phys.}, 15:125002, 2013.

\bibitem{Bassa2015}
H.~Bassa, S.K. Goyal, S.K. Choudhary, L.~Uys, H.~Di\'{o}si, and T.~Konrad.
\newblock Process tomography via sequential measurements on a single quantum
  system.
\newblock {\em Phys. Rev. A}, 92:032102, 2015.

\bibitem{Jozsa1994}
Richard Jozsa.
\newblock Fidelity for mixed quantum states.
\newblock {\em Journal of Modern Optics}, 41(12):2315--2323, 1994.

\bibitem{Hiller2012}
M.~Hiller, M.~Rehn, F.~Petruccione, A.~Buchleitner, and T.~Konrad.
\newblock Unsharp continuous measurement of a bose-einstein condensate: Full
  quantum state estimation and the transition to classicality.
\newblock {\em Phys. Rev. A}, 86:033624, 2012.

\bibitem{AudretschDiosiKonrad02}
J.~Audretsch, L.~Di{\'o}si, and Th. Konrad.
\newblock Evolution of a qubit under the influence of a succession of weak
  measurements.
\newblock {\em Phys. Rev. A}, 66:022310, 2002.

\end{thebibliography}

\end{document}